\documentclass[12pt,amsmath,amssymb,aps]{revtex4-1}

\usepackage[utf8x]{inputenc}
\usepackage{ucs}
\usepackage{amsmath}
\usepackage{amsfonts}
\usepackage{amssymb}
\usepackage{graphicx}
\usepackage{natbib}
\usepackage{multirow}

\newcommand{\figWidth}{0.95}

\begin{document}
	\title{Sorting of capsules according to their stiffness: from principle to application}
	
	\author{Edgar H\"aner$^1$, Doriane Vesperini$^2$, Anne-Virginie Salsac$^2$, Anne Le Goff$^2$ and Anne Juel$^1$}
	\email{anne.juel@manchester.ac.uk}
	\affiliation{$^1$MCND and Department of Physics and Astronomy, University of Manchester, Oxford Road, Manchester M13 9PL, United Kingdom.\\ $^2$Biom\'{e}canique et Bioing\'{e}nierie, UMR CNRS 7338, Universit\'{e} de Technologie de Compi\`{e}gne, France.
	}
	
	\date{\today}
	
	\begin{abstract}
	We assess experimentally the ability of a simple flow-based sorting device, recently proposed numerically by [Zhu \textit{et al., Soft Matter}, 2014, \textbf{10}, 7705--7711], to separate capsules according to their stiffness. The device consists of a single pillar with a half-cylinder cross-section which partially obstructs a flow channel so that initially centred, propagating capsules deform and circumvent the obstacle into an expanding channel (or diffuser). We perform experiments with millimetric capsules of fixed size which indicate that the deviation of the capsule in the diffuser varies monotonically with a capillary number -- the ratio of viscous to elastic stresses -- where the elastic stresses are measured independently to include the effects of pre-inflation, membrane thickness and material properties. We find that soft capsules with resistance to deformation differing by a factor of 1.5 can be reliably separated in the diffuser but that experimental variability increases significantly with capsule stiffness. We extend the study to populations of microcapsules with size polydispersity. We find that the combined effects of increasing capsule deformability and relative constriction of the device with increasing capsule size enable the tuning of the imposed flow so that capsules can be separated based on their shear modulus but irrespectively of their size. 
	\end{abstract}
	
\maketitle
\section{Introduction}

The sorting of cells from complex suspensions such as blood is key to many medical diagnostic and treatment strategies \cite{Lim2014}. The propensity of cells to deform under mechanical loading is a biophysical marker which can help distinguish normal and cancerous cells. Cancerous cells are usually softer than their healthy analogues, but they may also be stiffer in a few instances. In fact, normal to cancerous cell stiffness ratios between 0.7 and 32 have been measured by atomic force microscopy in a wide range of human tissues \cite{Lekka2016}. In malaria, the progressive reduction in red blood cell (RBC) deformability is associated with the growth and development of the parasite inside the cell as well as an approximately threefold increase in shear modulus of the encapsulating membrane, measured locally with pipette aspiration. These effects have been shown to result in an approximately tenfold increase in apparent shear modulus of the RBC  \cite{Hosseini2012}. RBC disorders such as sickle cell disease are also associated with increasing cell stiffness -- an approximately threefold increase in membrane shear modulus \cite{Byun2012} -- and a broadening of the cell size distribution \cite{Guruprasad2018}. 

RBCs are examples of natural capsules consisting of a fluid core encapsulated by a semi-permeable membrane \cite{BarthesBiesel2016}. In fact, the transport and flow-induced deformation of RBCs and other biological cells are commonly modelled with idealised capsules and soft beads \cite{Pozrikidis03, Villone2017}. The complex fluid-structure interaction flows of single particles or suspensions depend of the relative magnitude of hydrodynamic and elastic stresses. Flow-based separation of cells and capsules according to stiffness generally relies on channel geometries which promote extensional flows and thus capsule deformation. However, the resistance to deformation of a capsule is sensitive to both shear modulus of the membrane and capsule size if assuming constant membrane thickness. This means that robust sorting devices must either pre-sort cells according to size or be insensitive to size polydispersity. In this paper, we demonstrate experimentally the sorting of capsules according to their shear modulus irrespectively of their size. 

Methods developed for the size-based separation of rigid, spherical particles such as deterministic lateral displacement (DLD) devices consisting of arrays of pillars \cite{HuangEtAl04, LongEtAl08} have been extended to sort particles according to their stiffness, as recently demonstrated with a two-dimensional numerical model of vesicle transport in a DLD \cite{Biros2019}. In contrast, inertial separation techniques \cite{DiCarlo09} have shown reduced efficiency when applied to deformable particles \cite{HurEtAl11}. The use of branched geometries has been investigated numerically in order to design a deformability-based sorting device  \cite{Villone2017} and explore the path selection of a capsule in the presence of inertia with a view to designing an enrichment device \cite{Wang2016,Wang2018}. Moreover, sorting according to deformability was recently demonstrated experimentally in a T-junction for capsules of fixed size \cite{Haener2020}. Moderate to high thoughput experimental cell-sorting methods are also emerging on microfluidic platforms. These include the deformation of single cells through generations of tapered constrictions driven by oscillatory flow \cite{Guo2016}, the flow through a periodic array of ridges oriented diagonally to the main stream \cite{Wang2013} and rapid image analysis of individual cells deformed in a cross-flow with a throughput of up to 2000 cells per second \cite{Gossett2012}. 

In this paper, we focus on a simple sorting device recently introduced numerically by Zhu et al. \cite{Zhu2014} where a single pillar, with a half-cylinder cross-section, partially obstructs a flow channel; see figure \ref{obstacle}. When capsules flowing along the centreline of the channel reach the pillar they are compressed along the flow direction and thus elongate tangentially to the curved obstacle. The deformation of the capsule as it circumvents the obstacle determines its trajectory in the downstream expanding channel (or diffuser) which it reaches after clearing the constriction created by the pillar. In the limit of Stokes flow, the deformation of the capsule is governed by an elastic capillary number, which corresponds to the ratio of viscous to elastic shear stresses. Zhu et al. \cite{Zhu2014} used three-dimensional boundary integral simulations to show distinct paths taken by capsules of different elasticity when circumventing the pillar. A microfluidic realisation of this device was used to demonstrate experimentally that elastic microcapsules differing by a factor of approximately three in shear modulus could travel along separate trajectories if the imposed flow was sufficiently strong \cite{VesperiniEtAl17}.

The objective of this paper is to comprehensively assess the sorting ability of the half-cylinder device for capsules of fixed size and different stiffness but also for polydisperse capsule populations. Experiments performed on neutrally buoyant, millimetric capsules of controlled shape, size and stiffness are used to demonstrate that the trajectories of constant-size capsules are governed by an effective capillary number based on a measure of the force required to deform the capsule between parallel plates. This metric accounts for the effect of pre-inflation and different membrane thickness in addition to variations in the elastic properties of the membrane. Hence, they can be reliably separated but we find that the device performance decreases significantly with increasing capsule stiffness. We then turn to the sorting of two populations of microcapsules with shear moduli differing by a factor of three, where each population comprises a wide range of capsule sizes between 20~$\mu$m and 120~$\mu$m. In addition to the increased propensity of the larger capsules to deform, the relative channel constriction which must be cleared by the capsule to reach the diffuser increases as the capsule size increase because the size of the device is fixed. We find that these two effects combine to separate capsules according to stiffness irrespective of the capsule size, provided that the capsules are larger than the width of this constriction. 

The outline of the paper is as follows. The methods used to fabricate and test  capsules in similar millimetric and micrometric devices are detailed in \S \ref{MatMeth} and the flow field in the microdevice is mapped out with rigid tracer particles in \S \ref{Delta}. The paths of capsules in the two devices are compared in \S \ref{ResTraj} for similar relative capsule sizes. The results of experiments on millimetric capsules of fixed size and polydisperse microcapsule populations are then presented in \S\S \ref{ResMilli} and \ref{MicroSort}, respectively. The significance of the results in terms of stiffness-based sorting is discussed in \S \ref{Disc}.

\section{Materials and Methods}
\label{MatMeth}

\subsection{Capsule preparation and characterisation}
\subsubsection{Millimetric capsules: }
\label{millicapsules}

\indent The millimetric capsules consisted of a liquid core encapsulated in a cross-linked ovalbumin-alginate membrane \cite{Levy96}. Their preparation and characterisation have been previously described in detail \cite{Haener2020}. Briefly, we prepared spherical ovalbumin-alginate gel beads by dropwise addition of a solution in water of sodium-alginate (1~\% w/v), propylene glycol alginate (2~\% w/v) and ovalbumin (8~\% w/v) to a solution of calcium chloride (10~\% w/v). We then cross-linked their outer-shell before re-liquefying the gel core of the bead. The manufactured capsules were stored in a saline solution (11 g/l NaCl) and reached equilibrium after approximately 24 hours. During this period, water permeated through their membrane, which resulted in the inflation of the capsules to a size larger than the initial bead radius by up to 25~\%, depending on membrane elasticity and initial size. This implies the presence of a significant pre-stress in the capsule membrane. Solid beads were made in a similar manner to capsules, except that their core was not re-liquefied.

We selected four capsules with an average diameter $D= 3.90 \pm 0.03$~mm and one elastic bead $D_\textrm{bead}=3.83$~mm for experimentation \cite{Haener2020}. The capsules were from several batches manufactured under different experimental conditions in order to access different stiffness values. Hence, the capsules also had different values of inflation ($1.15 \le D/D_{\rm bead} \le 1.24$), membrane thickness $h$ ($0.18 \le 2h/D \le 0.23$) and sphericity (ratio of the minimum to maximum diameter $0.88 \le D_{\rm min}/D_{\rm max} \le 0.93$). The sphericity of the elastic bead was $D_{\rm min}/D_{\rm max}=0.85$. 

We characterised the elastic properties of each capsule by measuring the constitutive relation that governs capsule deformation by compression testing between parallel plates. An Instron 3345 Single Column Testing System (5 N load cell, accuracy $\pm$0.5 mN) was used to measure the force exerted by a top plate lowered quasi-statically to compress a capsule placed on an anvil within a saline bath \cite{Haener2020}. Measurements were performed at most three days before conducting the flow experiments since the age of the capsule influences its mechanical properties significantly. The large deformations routinely observed when capsules were propagated in flow meant that a nonlinear form of the constitutive law was necessary to characterise their deformation. In addition, their wall thickness was significant which meant that a membrane model was not appropriate for these capsules and the stiffness due to pre-inflation is not captured by the surface shear modulus of the capsule membrane. Hence, we chose the force required to deform a capsule to 50\% of its original diameter, $F_{50\%}$, as a direct experimental measurement of the capsule resistance to deformation. The measured values of $F_{50\%}$  for each capsule and the elastic bead are listed in Table \ref{table:Capsules}.

\begin{table}[h]
	
	\centering
	
	\begin{tabular}{c|cccc|c}
		&  \multicolumn{4}{ c| }{Capsules} & Elastic bead \\ 
		&  C1 \quad& C2 \quad& C3 \quad & C4 \quad& C5 \quad\\ \cline{1-6}\\ 
		$F_{50\%}$ & 4.5 & 6.2 & 7.4 & 21.9 & $> 33$\\ 
		$\pm 0.5$ (mN) &&&&&\\  \cline{1-6}
	\end{tabular}
	
	\caption{Force required to compress the selected capsules and the elastic bead to 50~\% of their initial height during compression tests in a saline bath using an  Instron 3345 Single Column Testing System (5 N load cell, accuracy $\pm$0.5 mN)\cite{Haener2020, Haener2017}.}
	\label{table:Capsules}
\end{table}

\subsubsection{Micrometric capsules: }

Ovalbumin microcapsules were prepared by interfacial cross-linking \cite{Andry1996}. Briefly, a water-in-oil emulsion was formed by mechanical agitation using a 10~\% ovalbumin solution in a phosphate buffer at pH 5.9 or pH 8 for the preparation of soft and stiff capsules, respectively. Cyclohexane added with 2~\% (w/v) sorbitan trioleate was used as the external phase. 
Cross-linking of ovalbumin at the oil/water interface was induced by adding 2.5~\% (w/v) terephtaloyl chloride to the organic phase. After 5 minutes, the chemical reaction was stopped by dilution with a solution of chloroform:cyclohexane (1:4, v/v). Capsules were then rinsed with an aqueous solution of polysorbate, followed by pure water. They were stored in water at 4$^\circ$C and suspended in glycerol when needed for the flow experiments.
The viscosity of the encapsulated fluid (supernatant of the capsule suspension) was measured using a cone-plate viscometer (Haake) and found to vary in the range $0.8 \le \mu \le 0.9$~Pa~s between experiments. 


The elastic properties of the microcapsules were determined by inverse analysis techniques \cite{Chu2011}. Capsules were imaged during steady propagation in a straight, fluid-filled capillary tube of circular cross-section to measure their two-dimensional contours in the mid-plane and velocity $U$. The viscosity $\mu$ of the suspending fluid was measured separately. The shape of an initially spherical capsule flowing through a narrow capillary of diameter $d$ depends solely on two parameters: the confinement ratio $D/d$, which can be extracted from experimental data, and the capillary number, defined as the ratio of viscous to elastic forces $Ca = \mu U/G_s$, where $G_s$ is the surface shear modulus of the capsule and $U$ is the mean velocity of the suspending fluid \cite{Lefebvre2008}. The capillary number, and thus the value of $G_s$, were determined for each capsule by comparison between the contour shape and a library of numerical profiles obtained for the appropriate confinement ratio and a range of $Ca$ \cite{Chu2011, Hu2013}. Values of the surface shear modulus $G_s = 0.030 \pm 0.007$~N/m and $G_s = 0.081 \pm 0.026$~N/m were obtained for the capsules fabricated at pH~5.9 and pH~8, respectively, with 5 min reticulation time. These results are consistent with previously published characterisation of capsules prepared according to the same protocol \cite{Chu2011}.

The polydispersity of the two microcapsule populations was quantified in sample of approximately 170 
capsules, which were confined between a glass slide and a cover slip separated by a 150 $\mu$m thick spacer. Images from an inverted microscope (DMIL LED, Leica Microsystems GmbH, Germany) were analysed with ImageJ to determine the diameter of each capsule using a circular fit.
We found that the mean and standard deviation of the capsule diameter to be $D = 60 \pm 18 \ \mu \mathrm{m}$ and $D=79 \pm 21\ \mu \mathrm{m}$ for stiff and soft capsules, respectively. 

\begin{figure}
	\centering
	\includegraphics[width= \figWidth \textwidth]{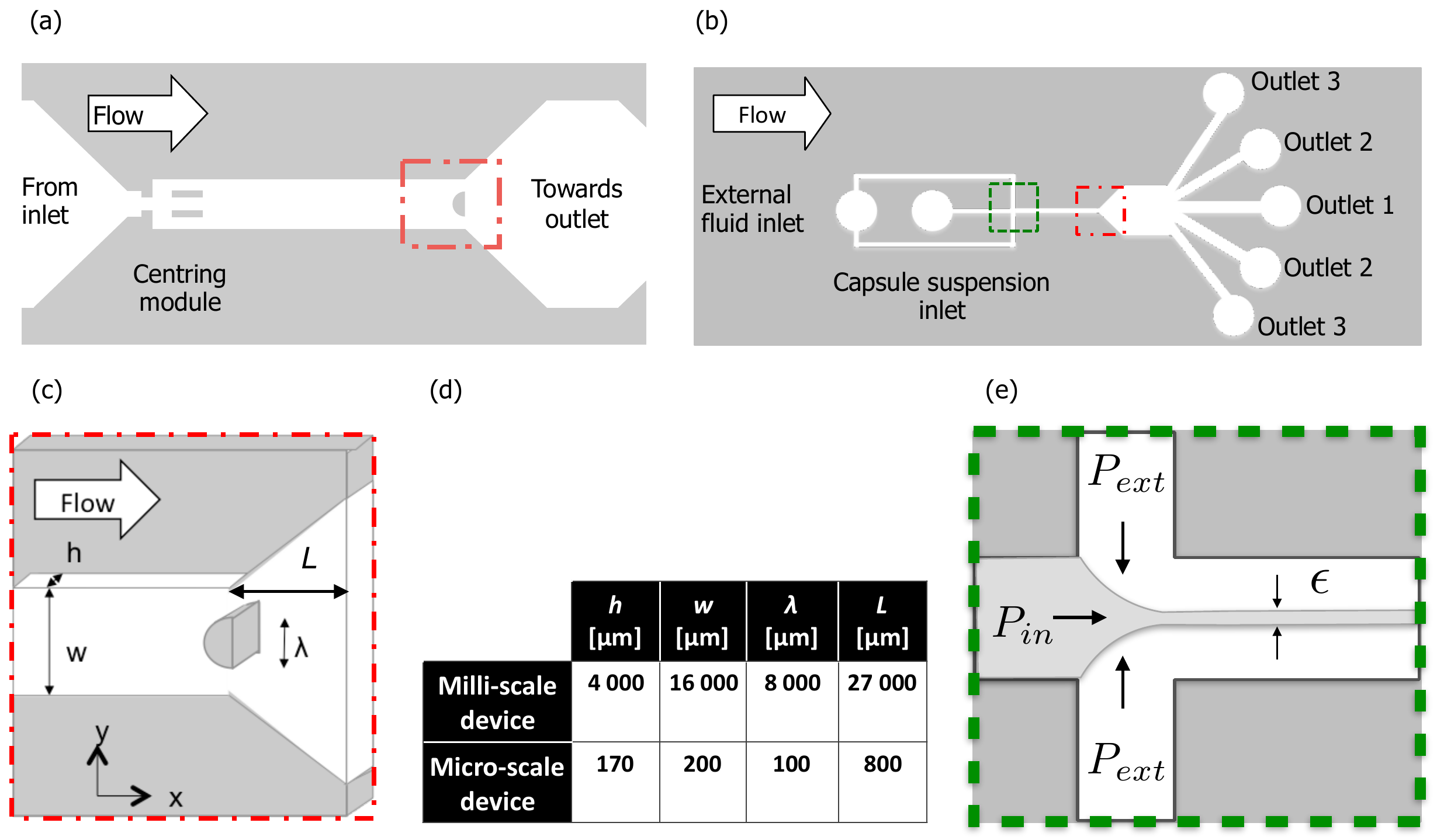}
	\caption {Schematic diagram of the sorting device viewed from above. Fluid and capsules flow from left to right. (a) Millifluidic device. The centring module, which is aligned along the centreline of the channel, consists of two parallel beams which are 10~mm long and separated by a distance of 4~mm. (b) Microfluidic device. (c) Close-up view of the half-cylinder obstacle of diameter $\lambda$. The capsules propagate in the rectangular channel of width $w$ and height $h$ with an offset $\kappa$ of their centroid from the centreline of the channel. Both geometries include centring modules to minimize $\kappa$. The capsules deform when they approach the obstacle, retain their deformation while circumventing it and relax their shape as they flow into the diffuser of length $L$ and opening angle of $45^\circ$ which leads to the outlets. 
		(d)  Comparison between the key dimensions in the millifluidic and microfluidic devices. (e) Schematic diagram of the flow-focusing module used to perfuse and centre the capsules in the microscale device. Capsules suspended in glycerol were injected at $P_{\rm in}$ and this centred fluid thread was narrowed by injection of pure glycerol at $P_{\rm ext}$ through the cross-branches of the module to a width $\epsilon < D$. 
	}
	\label{obstacle}
\end{figure}   

\subsection{Sorting devices}
\label{sorting}

Similar sorting devices were used to separate millimetric and micrometric capsules according to their stiffness 
(figure \ref{obstacle}(a,b)). They consisted of a main channel of rectangular cross-section with width $w$ and depth $h$ listed in figure \ref{obstacle}(d) alongside other key dimensions. In the microfluidic device, the capsule confinement ratio varied with the diameter of the capsule ($0.10 < D/w < 0.59$ and  $0.12 < D/h < 0.71$), whereas it was fixed in the millifluidic experiments because of the approximately constant capsule diameter ($D/w = 0.24$ and $D/h = 0.97$). The main channel was partially obstructed at its downstream end by a standing half-cylinder of diameter $\lambda$ and height equal to the channel depth, with its curved surface facing the main channel, as shown in figure \ref{obstacle}(c). Both devices comprised centring devices in order to minimise the offset $\kappa$ of the capsule centroid from the centreline of the main channel. Past the obstacle, the width of the channel increased linearly so that the stiffness-dependent displacement of the capsule imposed by the obstacle could be amplified. We henceforth refer to this region of the device as the diffuser, which featured an opening angle of $45^\circ$ in both devices. 
We now describe the experimental details pertaining to each sorting device in \S\S \ref{milliexp} and \ref{microexp}. 
\subsubsection{Millifluidic experiments: }
\label{milliexp}

The millimetric sorting device (figure \ref{obstacle}(a)) was machined out of cast acrylic sheets (Perspex, Gilbert Curry Industrial Plastics Co Ltd.) using a CNC milling machine. It consisted of two facing plates screwed together. Each plate was milled flat to within 25~$\mu$m prior to the milling of the channels. The top plate featured fluid inlets and reservoirs while the millimetric channels were milled into the bottom plate. 

The sorting device was levelled horizontally to within $0.5^{\circ}$ and back-lit with a custom-made LED light box. Capsule propagation was recorded in top-view at frame rates between 1 and 250  frames per second, using a monochrome CMOS camera (PCO, 1200hs) and a 50 mm micro-lens (Nikon) mounted vertically above the experiment. The capsules were propagated in degassed silicone oil (polydimethylsiloxane, Dow Corning, 5000~cS) of viscosity $ \mu = 5.23 \pm 0.04$ kg~m$^{-1}$s$^{-1}$ and density $\rho= 970 \pm 10$ ~kg~m$^{-3}$ measured at the laboratory temperature of $21^{\circ} \pm 0.5^{\circ}$C. The capsules were approximately neutrally buoyant, with a downward vertical drift velocity of less than 0.2~mm~min$^{-1}$ when freely suspended in a beaker of silicone oil. 

Before inserting a capsule into the liquid-filled device, any water on its outer surface was removed with a paper towel. The difference between the refractive indices of water and silicone oil makes water easily identifiable in the experimental images. If a water film initially coated the capsule, it could be isolated and removed  from the experiment by pushing the capsule through the flow channel at a high flow rate. At the start of each experiment, a capsule was positioned at the inlet of the device. A constant flow was then imposed by injecting silicone oil into the inlet of the flow device using a syringe pump (KDS410, KD Scientific) fitted with a 50~ml stainless steel syringe (WZ-74044-36, Cole-Parmer). In all the flow experiments, the Reynolds number was $Re= \rho\, Q \ /(\mu h) <  9 \times 10^{-2}$. A centring device consisting of two 10~mm long parallel beams separated by a distance of 4~mm was positioned near the inlet of the device in order to minimise the offset of the capsule centroid to typical values $\kappa \le 0.5$~mm.

The visualisation in top-view yields images of the capsules in the horizontal plane of view averaged over the depth of field. The centroid of the projected area of the capsules was determined from the contours extracted from these images. We used Python 2.7 and OpenCV ({\tt https://opencv.org/}), a cross-platform, open-source computer vision library, for the image analysis. We applied background subtraction with either an adaptive threshold or a simple uniform threshold determined via the Otsu method \citep{Otsu79}. In the resulting black and white image, all contours were identified with a Canny filter \citep{Canny86}. 


\subsubsection{Microfluidic experiments: }
\label{microexp}
The microfluidic sorter (figure \ref{obstacle}(b)) was fabricated in polydimethylsiloxane (PDMS) using standard soft lithography methods.
Briefly, a mixture of liquid PDMS and crosslinker (Sylgard 184, Dow Corning) was poured onto a photoresist-coated silicon master (Microfactory, Paris), degassed and cured for 2 hours at 70$^\circ$C. 
Two inlets and five outlets were created using a 0.75~mm biopsy punch before the chip was sealed onto a glass slide following plasma treatment. PTFE tubing was used to connect the chip to fluid reservoirs. The length of the inlet (outlet) tubes was 80 cm (25 cm) and their inner diameters 0.3 mm (0.56 mm), respectively. The outlet reservoirs were open to the ambient air at atmospheric pressure.

In contrast with the millifluidic device, flow was driven by a constant pressure head imposed by a pressure controller (MFCS, Fluigent, France). Two pressurised reservoirs supplied fluid via the two inlets to a flow-focusing module sketched in figure \ref{obstacle}(e), which was located at the upstream end of the main channel. Injection of a suspension of capsules in glycerol from the first reservoir at pressure $P_{\rm in}$ led to a fluid thread which was narrowed to a width $\epsilon$ by streams of pure glycerol injected via the cross-branches of the flow-focusing module from the second reservoir at pressure $P_{\rm ext}$. We adjusted the values of $P_{\rm in}$ and $P_{\rm ext}$ so that $\epsilon < D$ which enabled the centring of the capsule in the main channel. We imposed two values of the pressure head with $(P_{\rm ext},P_{\rm int})=(1000\; \mathrm{mbar},\,200 - 267\; \mathrm{mbar})$ and $(P_{\rm ext},P_{\rm int})=(4500\; \mathrm{mbar},\, 1200\; \mathrm{mbar})$, which we will refer to henceforth as low and high flow strengths, respectively. These resulted in a wide distribution of offsets of the capsule centroid with an average value of $\kappa = 5\ \mu \mathrm{m}$ for soft capsules and stiff capsules at low flow strength. Offsets were larger for stiff capsules at high flow strength with an average value of $\kappa = 9\ \mu \mathrm{m}$ \cite{Bihi2019}. 
 We measured the capsule velocity upstream of the 
obstacle for both flow strengths and capsule populations. For soft capsules, we found $U=1.0 \pm 0.2$~mm/s at low flow strength, and $U=5.0 \pm 0.3$~mm/s at high flow strength and for stiff capsules, $U=0.7\pm0.1$~mm/s at low flow strength, and $U=4.8 \pm0.6$~mm/s at high  flow strength.

During operation, the sorter was placed on the stage of an inverted optical microscope (DMIL LED, Leica Microsystems GmbH, Germany) equipped with a high speed camera (Fastcam SA3, Photron, USA).
Images were acquired at 5000 frames per second. A $10 \times$ magnification was chosen so that propagating capsules could be tracked from the obstacle to the diffuser. Images were analysed using the ImageJ software.

\subsection {Flow field in the microfluidic sorter}
\label{Delta}

The role of the diffuser is to amplify the stiffness-dependent displacement of the capsule imposed by the obstacle by increasing the separation between capsules of different stiffness.
Prior to introducing capsules to the device, we mapped typical trajectories past the obstacle and through the diffuser by propagating Lycopodium spores (Sigma Aldrich). These are rigid particles with a diameter of approximately $20 \ \mu \mathrm{m}$, which is less than half the width of the narrowest passages in the device between the half-cylinder and the side walls of the main channel, hereafter referred to as the constriction. Therefore, they could be made to propagate through the narrow passages at different distances from the obstacle.  

\begin{figure}[!ht]
	\centering
	\includegraphics[width= \figWidth \textwidth]{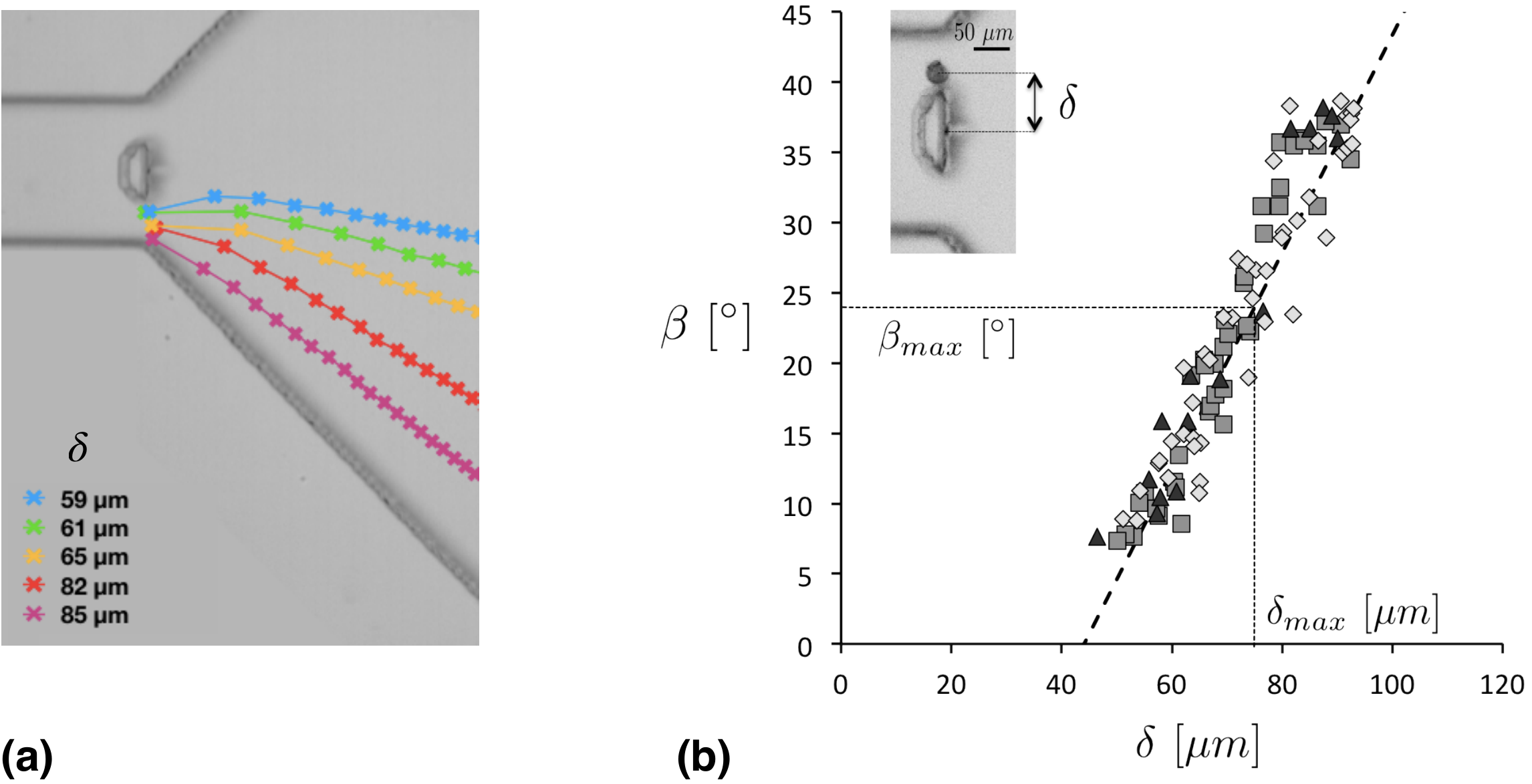}
	\caption {(a) Trajectories of lycopodium particles with a diameter of approximately $20\;\mu$m measured under high flow strength, for different values of the distance $\delta$ between the particle centroid and the channel centreline, defined graphically in the inset of (b). (b) Variation of the trajectory angle $\beta$ as a function of $\delta$ for lycopodium particles. Grey squares, triangles and diamonds correspond, respectively, to $P_{in} = 267$, 1200, 2400 {\ }\mbox{mbar} and $P_{ext} = 1000$, 4500, 6000 {\ }\mbox{mbar}. The dashed line is a linear fit to the data. The distance $\delta_{\rm max}= 75\;\mu$m corresponds to the distance from the channel centre line of the centroid of a non-deformable object of diameter $D= 50\;\mu$m (width of the constriction) in this device. The angle $\beta_{\rm max}=24^\circ$ is the trajectory angle associated with $\delta_{\rm max}$.}
	\label{lyco}
\end{figure}

By injecting a very dilute suspension of Lycopodium powder in glycerol through the cross-branches of the flow focusing module (figure \ref{obstacle}(e)) at ($P_{\rm ext}, P_{\rm int}) = (4500\; \mathrm{mbar},1200\; \mathrm{mbar})$, we captured a range of different trajectories, which are shown in figure \ref{lyco}(a). The experimental pathlines are reminiscent of Stokes flow in a diverging channel although streamlines are locally distorted by the finite-size particle. This is consistent with the small value of the Reynolds number based on the obstacle size $Re = \rho V \lambda / \mu \sim 10^{-4}$. 
The distance between two neighbouring trajectories increases monotonically as particles travel downstream of the obstacle and their pathlines become approximately linear in the second half of the diffuser. We quantified the separation between particles by measuring the displacement $\delta$ of the particle centroid from the centre line of the channel (see the inset of figure \ref{lyco}(b)) and the angle $\beta$ between the linear portion of the trajectory in the diffuser and the channel centre line.

Figure \ref{lyco}(b) shows that $\beta$ increases approximately linearly with $\delta$. Each symbol indicates a different value of the driving pressure $P_{\rm ext}$, taking care to adjust the internal pressure $P_{\rm in}$ in order to ensure a thin thread of inner fluid. All the data collapse onto a master curve fitted by a straight line. It indicates that the trajectories are independent of the flow strength within the parameter range explored, thus confirming the absence of inertial effects.
The limits of $\beta =0$ (trajectory parallel to the channel centre line) and $\beta = 45^\circ$ (trajectory adjacent to the diffuser walls) cannot be reached because of the finite size of the particles.
The trajectory angle of a particle which flows through the centre of the constriction ($\delta_{\rm max} = \lambda/2 + (w-\lambda)/4 = 75 \ \mu \mbox{m}$) is approximately $\beta_{\rm max}=24 ^\circ$ in the microfluidic device. Note that we expect a similar value in the millifluidic device because the size ratio $w/\lambda$ is the same and so is the opening angle of the diffuser. Values of $\beta > \beta_{\rm max}$ correspond to particles travelling through the constriction closer to the channel wall than to the obstacle wall. When capsules are propagated along the centre line of the main channel, as discussed in \S \ref{microexp}, they deform as they approach the obstacle and circumvent it adjacent to its surface. Also, the capsules that flow past the obstacle necessarily have a local width that is less than the constriction width $(w-\lambda)/2$. Hence, they are associated with a centroid position $\delta \le \delta_{\rm max}$, which yields $\beta \le \beta_{\rm max}$.


\section{Results}
\label{results}

\subsection{Capsule trajectories through the sorting devices} 
\label{ResTraj}

\begin{figure}[hp!]
	\centering
	\includegraphics[width=0.7 \textwidth]{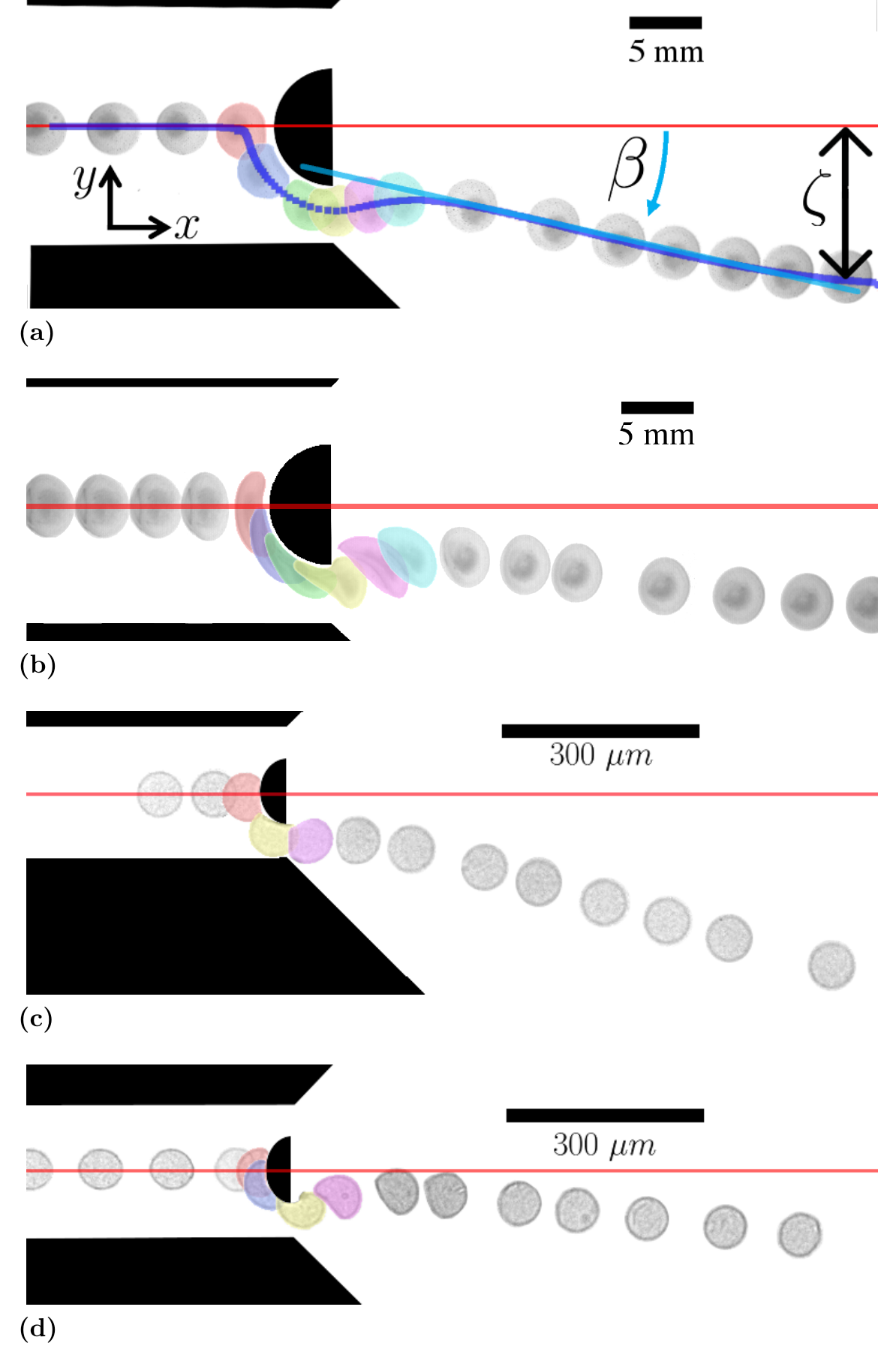}
	\caption{Sequences of time-lapse experimental images showing the propagation of a single capsule through the sorting device. 
		(a,b) Capsule C1 in the millifluidic device: (a) $Q = 2$~ml/min;
		(b) $Q = 50$~ml/min.
		(c,d) Microfluidic device with soft capsules of diameter $D=60$~$\mu$m:	(c) low flow strength $(P_{\rm ext},P_{\rm in}) = (1000\; \mathrm{mbar},200-267\; \mathrm{mbar})$; (d) high flow strength $(P_{\rm ext},P_{\rm in}) = (4500\; \mathrm{mbar},1200\; \mathrm{mbar})$. In each panel, the centre line of the channel is shown with a red, solid line. In (a), the trajectory of the capsule centroid is shown with blue squares. A solid blue line indicates the portion of diffuser where the capsule trajectory is approximately linear. $\zeta$ is the distance of the capsule centroid from the centre line at the end of the diffuser. $\beta$ is the angular deviation of the capsule trajectory from the centerline.}
	\label{FigTraj}
\end{figure}

Typical capsule trajectories are shown in figure \ref{FigTraj}, where time-lapse images indicate how a single capsule travels from left to right through the millifluidic device (a, b) and the microfluidic device (c, d). The time interval between images varies depending on the dynamics of the capsule. In each panel, the capsule initially propagates with constant velocity towards the half-cylinder obstacle along the centre line of the channel.
As the capsule approaches the obstacle, the imposed flow compresses the capsule along the streamwise direction so that it starts to elongate orthogonally to the centre line (see, e.g., figure \ref{FigTraj}(b)). This compression increases significantly when the distance from the centroid of the capsule to the apex of the obstacle falls below one capsule radius, and the capsule then deforms by elongating approximately tangentially to the obstacle. If the capsule were perfectly centred and both flow and capsule were perfectly symmetric about the centre line of the channel, the capsule would remain trapped at the stagnation point on the apex of the obstacle. Instead, the small offset $\kappa$ (see \S \ref{sorting}), unavoidable in experiments, allows the capsule to continue to travel past one side of the obstacle into the diffuser. The sign of $\kappa$ determines the side of the half-cylinder circumvented by the capsule. The frames where the capsule is adjacent to the obstacle have been coloured in order to highlight the evolving deformation of the capsule in this region.

The same millimetric capsule (C1) is shown for low and high values of the flow rate in figures \ref{FigTraj}(a,~b), while in figures \ref{FigTraj}(c,~d) two microcapsules of similar diameter are shown for low and high pressure heads, respectively. At low flow velocities (figures \ref{FigTraj}(a,~c)), the capsules deform only weakly on the obstacle and their centroid therefore follows streamlines into the diffuser that diverge significantly away from the centreline of the device. In contrast, the larger capsule deformations incurred at high flow velocities (figures \ref{FigTraj}(b,~d)) mean that the capsule centroids approach the surface of the obstacle more closely and thus that the capsules are entrained along streamlines that remain closer to the centre line. 


All the capsules propagate along approximately straight lines in the diffuser when they are more than two capsule diameters from its extremities (in the streamwise $x$-direction). This is illustrated in figure \ref{FigTraj}(a) where the trajectory of the capsule centroid is shown with blue squares and a superposed solid blue line highlights the region of the diffuser where the trajectory is linear. As in \S \ref{Delta}, we quantify the deviation of the capsules from the centre line 
with the angle $\beta$ spanning the arclength from the centre line of the device to the capsule trajectory, which is independent of the measurement location along the diffuser. We also define the capsule displacement $\zeta$ as the distance separating the centroid of the capsule from the centreline at the end of the diffuser. Figure \ref{FigTraj} shows that $\beta$ decreases for increasing flow rate or pressure head. 
The use of different types of capsules and different capsule confinement ratios in the two devices suggests that this behaviour is robust.

\subsection{Single capsule experiments}
\label{ResMilli}

\subsubsection{Influence of flow rate and capsule stiffness on capsule trajectories in the diffuser: }
\label{FlowRate}

\begin{figure}[hpt]
	\begin{tabular}{cc}
		(a) & (b) \\
		\includegraphics[width= 0.47\linewidth]{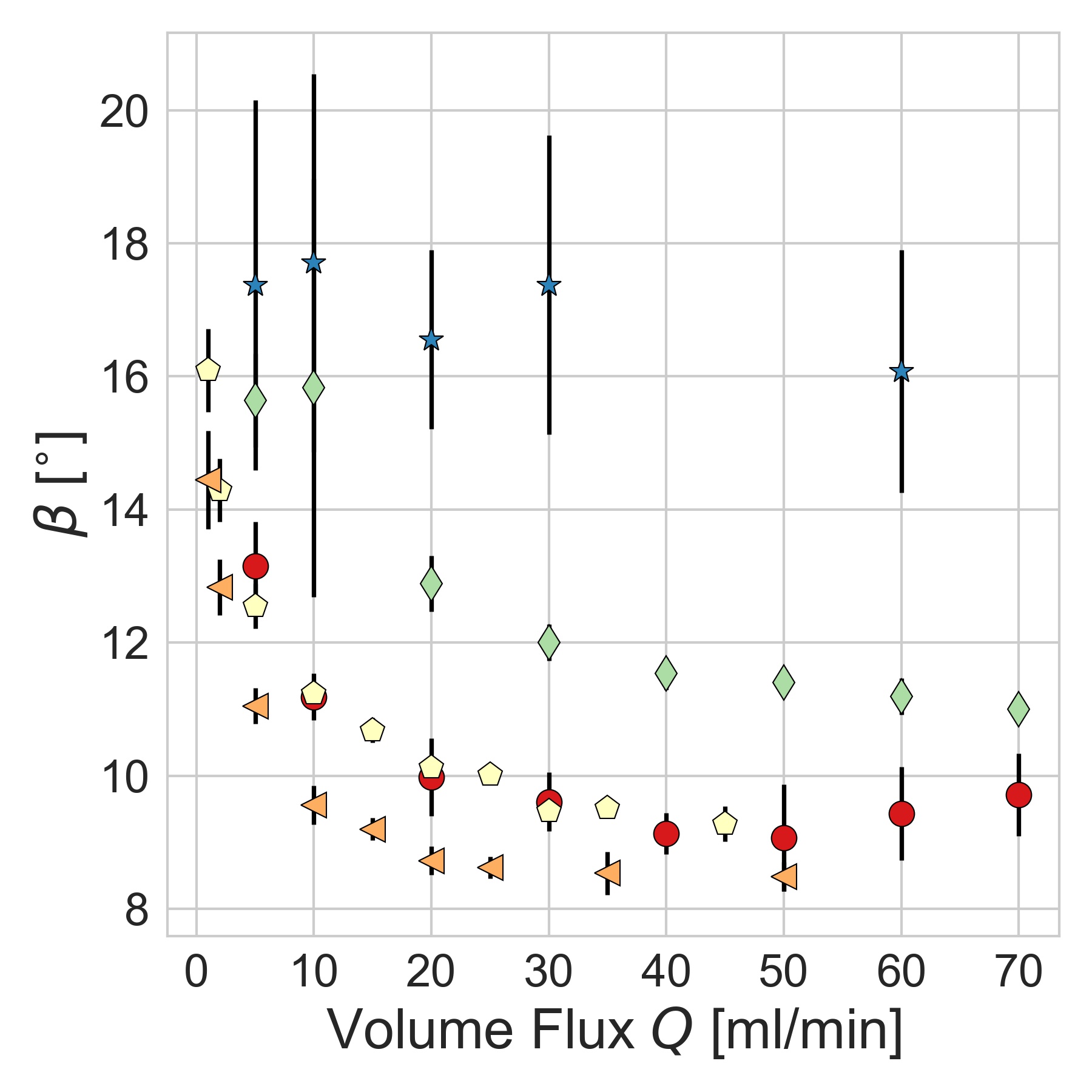}
		& \includegraphics[width= 0.47 \linewidth]{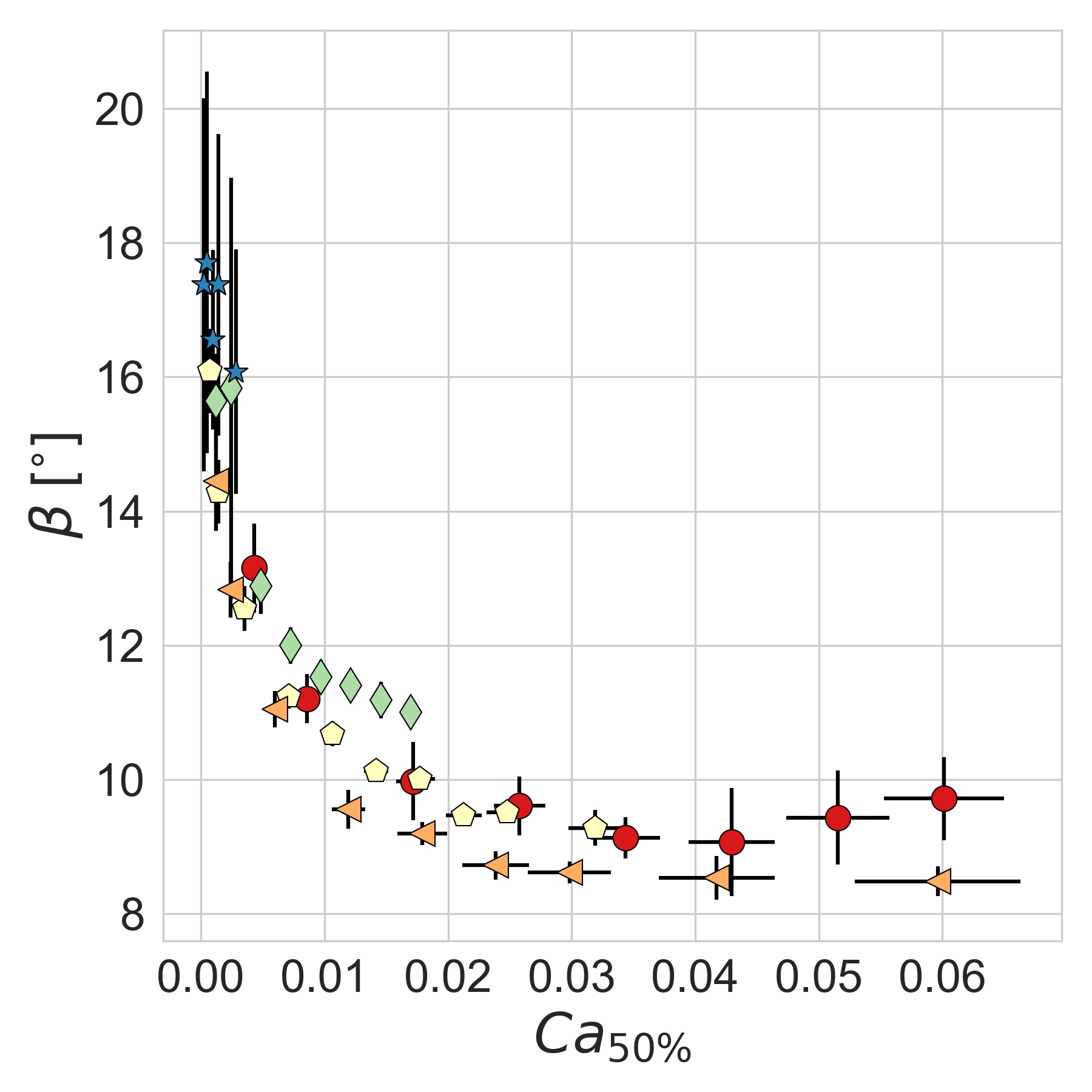}
	\end{tabular}
	\caption{Deflection angle of capsule trajectories in the millimetric half-cylinder device: (a) as a function of volume flux and (b) as a function of $Ca_{50\%}$. Orange triangles: capsule C1; red circles: C2; white pentagons: C3; green diamonds: C4; blue stars: elastic bead C5.  The symbols denote mean values, the vertical error bars indicate the standard deviation of multiple experiments and the horizontal error bars in (b) reflect the measurement error on the stiffness of the capsules. For the capsules, $|\kappa|$ \textless 0.5 mm ($|\kappa|/D \le 0.13$) and for the elastic bead C5, $|\kappa|$ \textless 1.2 mm ($|\kappa|/D \le 0.3$).}
	\label{FigFlow}
\end{figure} 

The trajectory angle $\beta$ is shown as a function of flow rate in figure \ref{FigFlow}(a) for millicapsules of different stiffness (C1--C4) and for the elastic bead C5 (see Table \ref{table:Capsules}). The symbols shown in figure \ref{FigFlow} correspond to mean values from four to fourteen experiments performed for each value of flow rate and the vertical error bars indicate the standard deviations. After each experiment, the capsule or elastic bead returned to its undeformed configuration and we did not detect any plastic deformation, consistent with the compression experiments discussed in section \ref{millicapsules}. For each capsule, $\beta$ decreases monotonically with increasing flow rate. This is because capsule deformation on the obstacle increases with $Q$ as previously shown in figure \ref{FigTraj}. For flow rates $Q \le 5$~ml/min, $\beta$ decreases more steeply the softer the capsule. This decrease becomes gentler for $Q \ge 5$~ml/min and a clear separation emerges between the trajectory angles of capsules of different stiffness, indicating that for a given flow rate $Q$, the trajectory angle $\beta$ increases with capsule stiffness. 

In contrast with capsule behaviour, the trajectory angle of the elastic bead (C5) remains approximately constant over the entire range of flow rates investigated. This is because the deformation of the elastic bead is small within this parameter range, so that its trajectory is approximately independent of the flow rate applied. Hence, the elastic bead, which is at least 50~\% stiffer than the stiffest capsule (C4) and 8 times stiffer than the softest capsule (C1), behaves approximately like a rigid sphere in this device. The trajectories of the elastic bead were therefore used to estimate the trajectory angle for a rigid particle, $\beta_{\rm max} = 17.6^{\circ} \pm 0.5^{\circ}$, which also corresponds to the maximum angle of a capsule in the limit of vanishing deformation. 
In contrast, the smallest trajectory angle recorded for the softest capsule (C1) is $\beta_{\rm min} = 8.6^{\circ} \pm 0.1^{\circ}$ for $Q \gtrsim 25$~ml/min. 

In figure \ref{FigFlow}(b), the trajectory angle is shown as a function of the elastic capillary number based on the force $F_{50\%}$ required to compress the capsule between parallel plates to half of its diameter (see section \ref{millicapsules}),
\begin{equation*}
Ca_{50\%} \ = \mu \frac{ Q}{hw} \frac{D}{F_{50\%} },
\label{eq:Ca}
\end{equation*}
where $\mu$ the viscosity of the fluid, $h w$ the cross-sectional area of the main channel and $D$ the capsule diameter \cite{Haener2020}. The main source of error on the capillary number is the $\pm 0.5$~mN uncertainty on the value on $F_{50\%}$, which is propagated into horizontal error bars in figure \ref{FigFlow}(b).  

The data collapse onto a monotonically decreasing master curve as a function of $Ca_{50\%}$ to within experimental uncertainty. After a steep initial decrease for $Ca_{50\%} \le 0.01$, $\beta$ appears to tend towards an approximately constant value at large capillary numbers ($Ca_{50\%} > 0.03$). The occurrence of a non-zero plateau value is expected in the limit of large $Ca_{50\%}$ (corresponding to infinitely soft objects) because the distance between the capsule centroid and the obstacle must remain greater than its membrane thickness in order to ensure the conservation of fluid within the capsule. However, we remain far from this idealised limit (see figure \ref{FigTraj}(b)), which would in practice be superseeded by capsule rupture. 

Figure \ref{FigTraj}(b) indicates that the capillary number is the only dimensionless parameter required to describe the displacement of capsules of fixed diameter in the half-cylinder millifluidic device. It demonstrates that the sorting strategy proposed by Zhu et al. \cite{Zhu2014}, in which the balance between hydrodynamic forces and capsule elasticity governs capsule deformation in the constriction and thus influences the trajectory, can be realised experimentally. Moreover, it could also enable the measurement of the relative stiffness of the capsules through experiments at a suitable constant flow rate that separates the capsules in the diffuser.

\subsubsection{Resolution of the sorting device: }
\label{Resolution}

\begin{figure}[ht]
	\centering
	\includegraphics[width=0.7\textwidth]{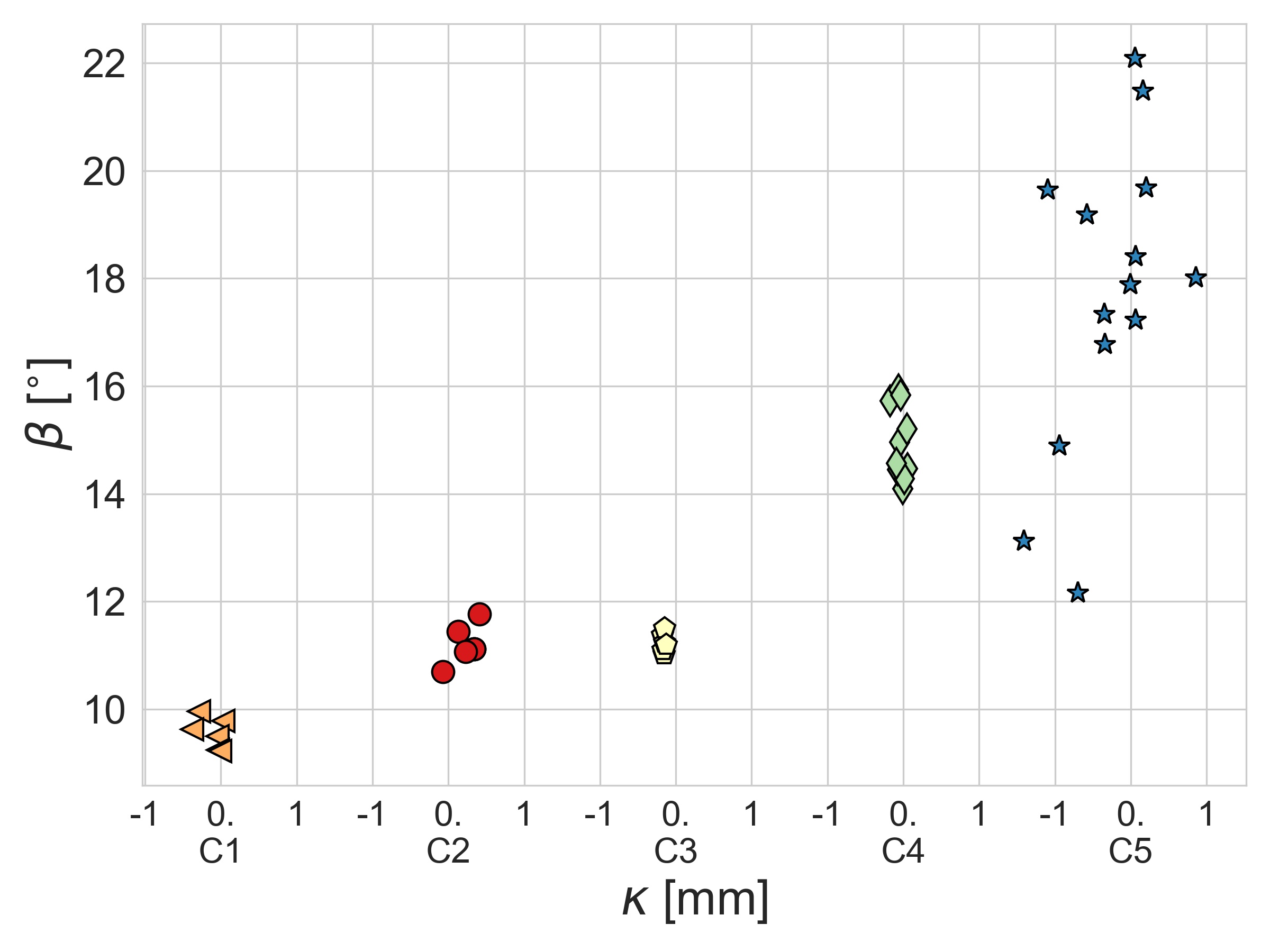}
	\caption{Trajectory angle $\beta$ from individual experiments plotted as a function of the offset $\kappa$ for each capsule shown in figure \ref{FigFlow} for $Q=10$~ml/min. The initial offset of the capsules is $ |\kappa| \le 0.5$ mm  ($|\kappa|/D \le 0.13$) and $ |\kappa| \le 1.2$ mm ($|\kappa|/D \le 0.3$) for the elastic bead C5.} 
	\label{FigBeta}
\end{figure}

The experimental measurements previously shown in figure \ref{FigFlow} demonstrate the separation on average of capsules of fixed size in the millifluidic device. However, in order to achieve reliable sorting, the trajectory angles of capsules with different stiffness must remain different in every repetition of the experiment. Figure \ref{FigBeta} shows all the values of $\beta$ obtained at a fixed flow rate of $Q=10$~ml/min. The data is grouped by particle and presented as a function of initial offset $\kappa$, using the same colour scheme as in figure \ref{FigFlow}. The initial offset of capsules, following their passage through the centring module (see figure \ref{obstacle}(a)) was $\mid\kappa\mid \, \le 0.5$~mm (i.e. $< 3~\%$ of the width of the channel) for the capsule experiments and $\mid\kappa\mid \, \le 1.2$~mm (i.e. $< 7.5~\%$ of the width of the channel) for the bead.

The flow behaviours of capsules C2 and C3, which have values of $F_{50\%}$ differing by less than 16~\%, are indistinguishable because of the experimental scatter of the trajectory angles of approximately 10~\%. 
Figure \ref{FigFlow} also shows that there is no overlap between the trajectory angles measured for capsules C1 and C2/C3, but that the largest value measured for C1 is only marginally smaller than for the smaller value obtained with C2. In fact, the minimum difference in trajectory angle between capsules C1 and C2 is $\Delta \beta = 0.8^{\circ}$,  and thus the minimum distance separating the centroids of the capsules at the outlet of the diffuser is $\Delta \zeta = 0.4$~mm (10 ~\% of diameter), in contrast with $\Delta \beta = 1.5^{\circ}$ ($\Delta \zeta = 1$~mm) on average. This means that the factor of approximately 1.5 between the stiffness of C1 and C2/C3 is the smallest contrast that can be resolved by the millimetric device. 

The most striking feature of figure \ref{FigBeta} is the significant increase in the experimental scatter of $\beta$ with increasing stiffness of the particle, which does not appear to be strongly correlated to the initial centring of the capsules, see also \S \ref{Disc}. The scatter varies from approximately $1^\circ$ for the three softest capsules to $2^\circ$ for capsule C4 and $10^\circ$ for the elastic bead C5. This means that the elastic bead cannot be reliably separated from capsule C4. However, capsule C4 can easily be separated from the softer capsules tested. We attribute this behaviour to the propensity of the softer particles tested to deform in flow (away from the obstacle). If the position of a capsule in a channel is perturbed, its shear-induced deformation generates lift forces that restore it to the centreline of the channel \cite{Doddi2008}. This means that the value of $\beta$ is only weakly sensitive to small positional errors and non-uniformities in the sphericity or membrane thickness of these softer capsules. In contrast, the absence of lift forces in the flow of elastic beads over the range of flow rates investigated renders their pathway highly sensitive to small variations in experimental conditions. Hence, our measurements indicate that the decreasing level of fluctuation in particle position as the particle becomes softer is a key factor in making sorting realizable in practice. 


Although capsules C1 and C2 are clearly separated in terms of the trajectory angle $\beta$, their actual separation distance measured at the outlet of the 27~mm diffuser amounts to $\sim 1 $ mm, which is less than the capsule diameter of 3.9~mm. Hence, the millimetric device provides a proof of concept of the sorting of capsules according to their stiffness, but segregating capsules into different channels would require a longer diffuser. This would enhance the separation distance at the cost of at least a tenfold increase in transit time, because a diffuser length $\ge 108$~mm would be required to achieve a separation of 4~mm.




\subsection {Sorting of capsule suspensions and effect of polydispersity}
\label{MicroSort}

Having demonstrated the sorting mechanism proposed by Zhu et al. \cite{Zhu2014} experimentally using carefully-manufactured individual millicapsules, we now address its application in microfluidic devices by investigating the flow of dilute microcapsule suspensions through the sorter. We focus on the influence of capsule polydispersity because the fixed size of the sorting device means that the confinement of the capsules as they propagate past the half-cylinder constriction depends on their size. Also, increasing the diameter of the capsules reduces their resistance to deformation, so that larger capsules are also more readily deformed when propagating through the sorter for given values of the flow strength and capsule shear modulus. 

\begin{figure}
	\centering
	\includegraphics[width= 0.95\textwidth]{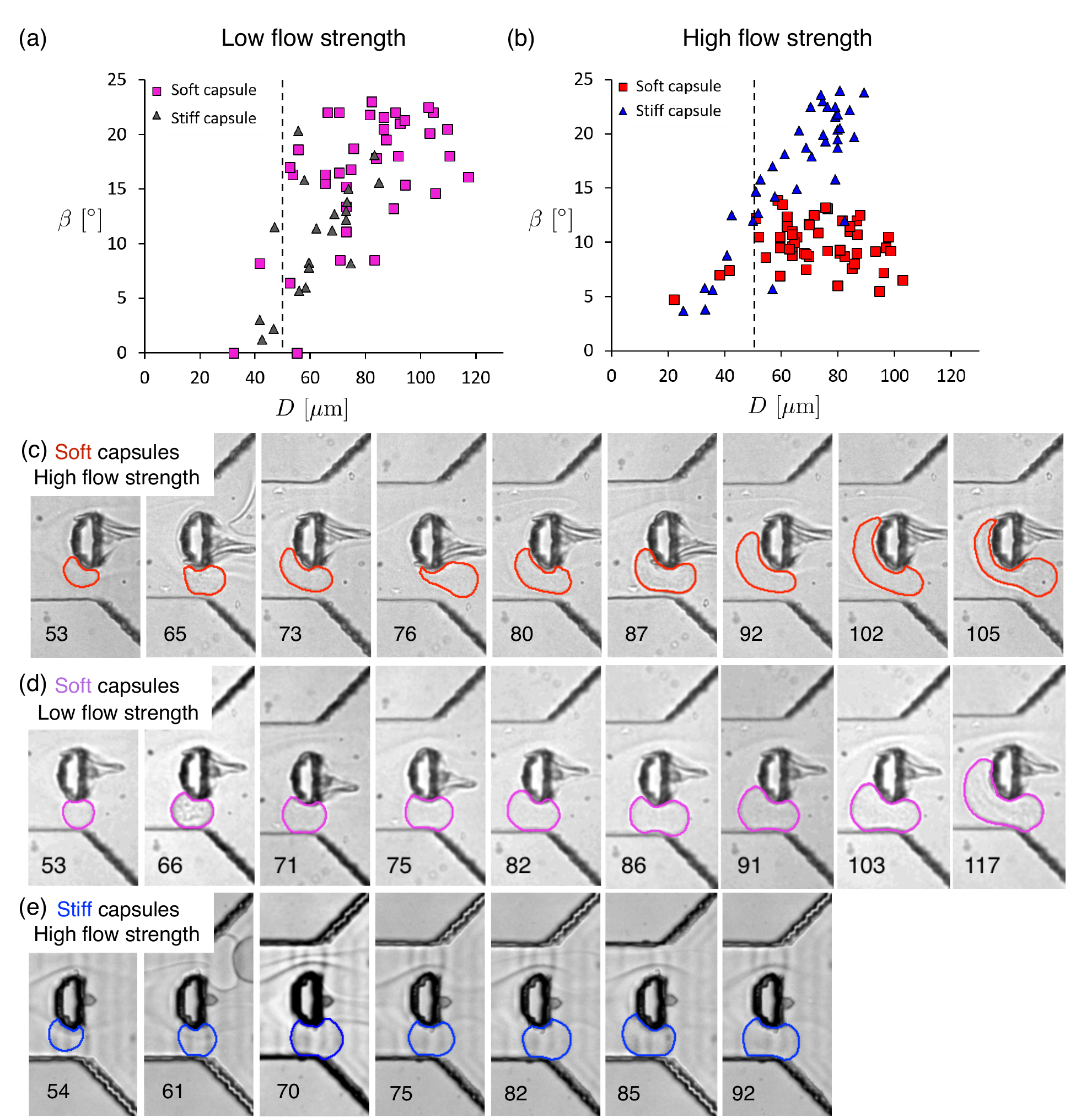}
	\caption{Trajectory angle $\beta$ as a function of the capsule diameter $D$: (a) low flow strength ($U = 0.9 \pm 0.2 {\ } \mbox{mm.s}^{-1}$); (b) high flow strength ($U = 4.9 \pm 0.4 {\ } \mbox{mm.s}^{-1}$). The maximum initial offsets $|\kappa|/D$ of the capsule centroids varied depending on capsule size between $0.1$ and $0.65$ for soft capsules and $0.16$ and $0.95$ for stiff capsules. Average values of $|\kappa|/D$ varied between $0.04$ and $0.25$ for soft capsules and $0.07$ and $0.45$ for stiff capsules. (c--e) Snapshots of the capsules in the most constricted section of the channel obstructed by the half-cylinder obstacle for the values of capsule diameter shown in (a,b): (c) soft capsules at high flow strength; (d) soft capsules at low flow strength; (e) stiff capsules at high flow strength. 
		The number in each image indicates the diameter of the capsule in $\mu$m. The small variations in the position of the capsules in the device at high flow strength is due to the limited number of frames available.}
	\label{FigSoft}
\end{figure}

Figure \ref{FigSoft}(a, b) compares the trajectory angles $\beta$ of soft and stiff capsules as a function of their diameter $D$. The dashed line at $D = 50 \ \mu \mathrm{m}$ indicates the minimum width of the constriction between the half cylinder and the channel wall. For low flow strength (figure \ref{FigSoft}(a)), the trajectory angles for soft and stiff capsules overlap across available capsule sizes and thus, the capsules are not separated in the sorter. By contrast, for high flow strength (figure \ref{FigSoft}(b)), the trajectory angles of soft and stiff capsules begin to diverge for $D \gtrapprox 50 \; \mathrm{\mu m}$. Despite significant scatter in the data, the increasing difference in trajectory angle with increasing diameter indicates that the sorter can reliably separate soft and stiff capsules with a stiffness factor of 2.7  for $D \gtrapprox 75 \; \mathrm{\mu m}$.

The millifluidic experiments reported in \S \ref{ResMilli} were performed for capsule diameters less than the constriction width and thus, they are most closely related to the microcapsule experiments performed for $D<50 \; \mathrm{\mu m}$. However,  the microsorter is unable to separate soft and stiff microcapsules within this capsule size range for both high and low flow strengths. This is because the pressure range available in the microfluidic device was not sufficient to impose flow strengths that significantly deformed the smaller capsules, even the soft ones. For example, soft capsules with diameters less than $D=32\; \mathrm{\mu m}$ did not visibly deform on the half-cylinder obstacle at high flow strength and likewise for diameters less than $D=53\; \mathrm{\mu m}$ at low flow strength. However, soft capsules exhibited increasing deformation as their diameter increased (see figure \ref{FigSoft}(c, d)). 

The trajectory angle of capsules with $D < 50\; \mathrm{\mu m}$ increases with $D$ in a similar way for soft and stiff capsules in figure \ref{FigSoft}(b) and this trend is also visible in figure \ref{FigSoft}(a), although the data is more scattered. This is because larger capsules occupy a larger proportion of the constriction so that their centroid positions move further away from the surface of the half cylinder resulting in larger trajectory angles. However, the trajectory angle does not exceed $24^\circ$ across experiments, consistent with the value $\beta_{max}$ identified in experiments with lycopodium particles, see \S \ref{Delta}. 

Capsules which exceed the width of the constriction ($D \ge 50 \; \mathrm{\mu m}$) even after deformation on the half cylinder obstacle are most strongly diverted from the centreline of the device because they fill the entire constriction as they squeeze through, see figure \ref{FigSoft}(d, e). This is the case for soft capsules at low flow strength (figure \ref{FigSoft}(a)) and stiff capsules at high flow strength (figure \ref{FigSoft}(b)). Their trajectory angle appears to saturate close to $\beta_\mathrm{max}$ as $D$ increases beyond the constriction width, albeit with significant scatter in the data. This is because the capsule front which first enters the diffuser when the capsule is pushed through the constriction remains approximately similar once the capsule is sufficiently elongated, i.e. for $D \gtrapprox  70 \; \mathrm{\mu m}$. There is also a size threshold beyond which capsules get trapped in the constriction, which depends on flow strength and stiffness. At low flow strength, the diameter of the largest stiff capsule to clear the constriction was $D=87 \; \mathrm{\mu m}$, while in the soft population, capsules up to $D=117\; \mathrm{\mu m}$ were found to propagate through the sorter. 

\begin{figure}
	\centering
	\includegraphics[width= 0.7\textwidth]{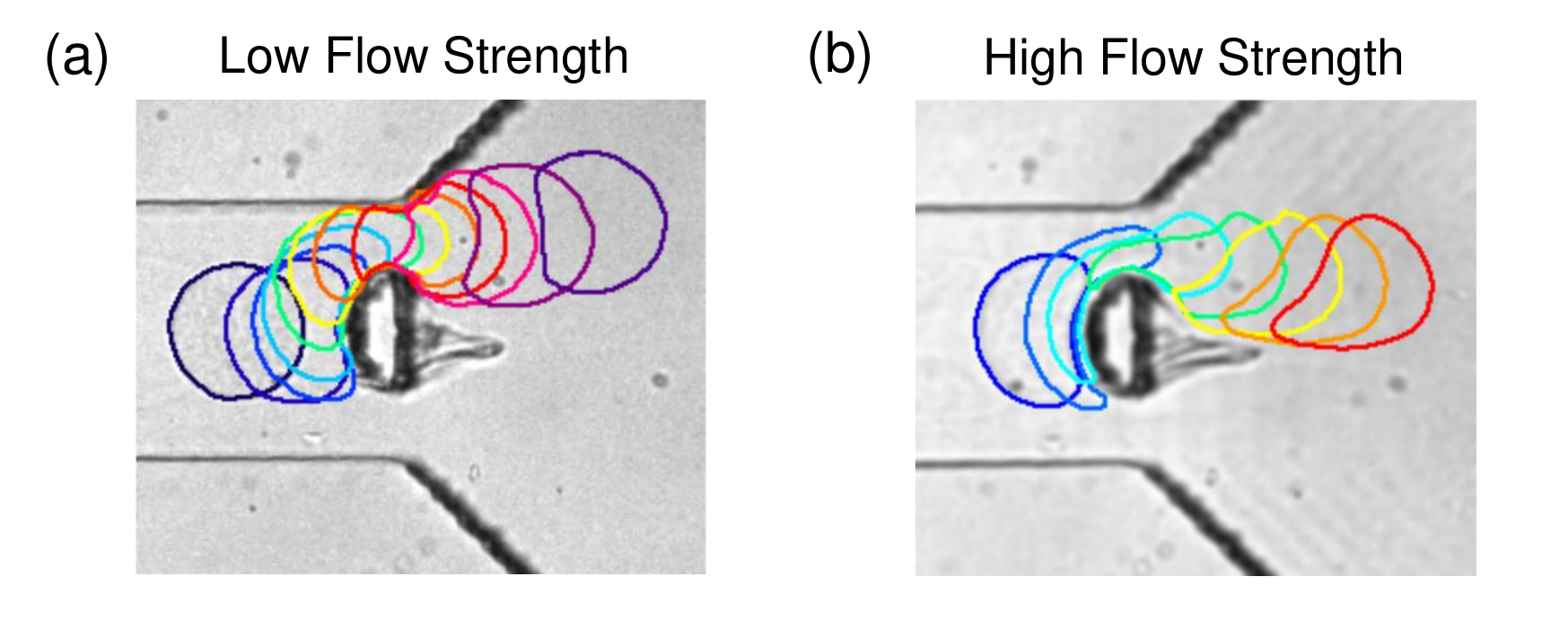}
	\caption{Time-lapse images of soft capsules with $D=103 \ \mu$m circumventing the half cylinder obstacle: (a) low flow strength; (b) high flow strength. The time interval between consecutive outlines is 20 ms at high flow strength, but variable at low flow strength. The duration of the sequence shown is 600~ms at low flow strength and 140 ms at high flow strength.
	The elongated grey trail behind the obstacle is due to particles (e.g., small fragments of broken capsules) which are trapped in the stagnation region behind the obstacle.}
	\label{FigLapse}
\end{figure}

Soft capsules at high flow strength are less strongly deflected by the half cylinder obstacle than at low flow strength (figure \ref{FigSoft}). Their trajectory angle appears to saturate at approximately $10^\circ$ for $D \gtrapprox 60 \; \mathrm{\mu m}$. The comparison between soft capsules circumventing the obstacle at low and high flow strength in figure \ref{FigLapse} indicates that the origin of the capsule deformation differs for high and low flow strengths. Whereas the constriction imposes the deformation of the capsule at low flow strength, at high flow strength the capsule elongates sufficiently on the obstacle so that it is narrower than the most constricted section of channel. Aided by the resulting shear-flow through the constriction, it appears to wrap around the obstacle so that its trajectory angle remains low. The increasing elongation of the capsule with increasing diameter means that the width of the capsule in the constriction remains approximately constant, resulting in a constant trajectory angle. Hence, the mechanism for separation of soft and stiff capsules at high flow strength is different from that proposed by Zhu et al. \cite{Zhu2014} in that it operates for capsules larger than the constriction and relies on the constriction to divert the stiffer capsules to large trajectory angles, while the softer capsules acquire low trajectory angles by deforming on the half cylinder obstacle in the usual way. 

\section{Discussion and conclusion}
\label{Disc}

We have tested the sorting device proposed in the numerical study by Zhu et al.  \cite{Zhu2014} with two complementary experimental approaches. The device consists of a half-cylinder centred at the end of a wide channel, so that a capsule propagating along the centreline of the channel is compressed onto the cylinder, resulting in its elongation tangentially to the cylinder's surface. The deformation of the capsule as it circumvents the obstruction determines its trajectory in the downstream, linearly-expanding channel. Thus, this device enables the sorting of capsules according to their stiffness. Using accurately manufactured millimetric capsules of fixed size and different stiffness, we have shown that the angle of deviation of the capsule trajectory from the centreline of the channel depends solely on an elastic capillary number, based on the force required to deform the capsule to 50\% in compression tests. This measure indicates the resistance to deformation of the capsule and thus, includes the effects of pre-stress and different membrane thickness which are significant in our millicapsules. We find that we can reliably separate capsules with a stiffness constrast down to a factor of 1.5, but that the resolution of the device degrades as the capsule stiffness increases. 

The second experiment used a microfluidic device to separate polydisperse populations of microcapsules according to their shear modulus. The fixed size of the device meant that the confinement of the capsules in the constriction increased with capsule size. However, their resistance to deformation decreases with size, so that larger capsules were more easily deformed. We demonstrated that two polydisperse populations of capsules with a factor of 2.7 between their shear moduli could be separated in the sorter provided that their diameter exceeds the width of the constriction by the side of the cylinder. This separation method relies on the stiff capsules filling the constriction as they squeeze through so that they adopt a large trajectory angle close to the maximum of $24^\circ$. In contrast, the soft capsules deform sufficiently on the obstacle so that they can adopt a low trajectory angle which remains constant at approximately $10^\circ$ because the width of the capsules as they circumvent the half-cylinder does not increase with capsule size. 

The large difference in trajectory angles achieved between the two populations is key to the robust sorting of these microcapsules because of the considerable fluctuations in the data from the microfluidic device. Although these may in part stem from imperfections in individual capsules, a key experimental parameter influencing the trajectory angle is the initial offset of the capsule from the centreline of the channel. In millicapsules, we found that for moderate offsets of up to $12.5\%$ of the capsule diameter variations in the trajectory angle $\beta$ (and therefore also the displacement $\zeta$) of up to $\pm 5\%$ were weakly correlated with $\kappa$, implying the influence of other experimental variabilities. Increasing the offset to $60\%$ of the capsule diameter yielded an approximately linear variation of the trajectory angle with offset, which increased by a factor of approximately 1.5. A similar trend was measured in the microfluidic device but the sensitivity of the capsule trajectories to the offset was considerably enhanced with an offset of $10 \%$ of the capsule diameter resulting in a $30\%$ increase in $\zeta$. Hence, the majority of the scatter in the data from the microfluidic device can be attributed to variations in initial capsule offset within that range. 

The ability to robustly sort polydisperse suspensions according to stiffness with a single device, without the need for sifting the particles according to size before perfusing them into the microchannel, opens new opportunities for high-throughput separation of cells, with applications to the diagnostic of, e.g., sickle cell disease, malaria or certain types of cancers. Our proof-of-concept experiments have shown applicability over parameter ranges relevant to cells, with the demonstration of sorting over a factor two in capsule size and a good sensitivity of a factor of  2.7 in shear modulus, both appropriate for cells.

\section{Conflicts of interest}
There are no conflicts to declare.

\section{Acknowledgements}
D.V. acknowledges financial support from the French Ministry of Research and E.H. was funded by an EPSRC DTP studentship. We acknowledge funding by the R\'egion Hauts-de-France (FORPLAQ project). We thank F. Edwards-Lévy (Université de Reims Champagne--Ardennes) for providing the microcapsules.





\footnotesize{ \bibliography{Capsules_Anne2}

\end{document}